\documentstyle[gcdv,psfig]{article}
\def\reference{\parskip 0pt\par\noindent\hangindent 0.5 truecm}
\def\kms{km ${\rm s}^{-1}$}
\def\lesssim{{_ <\atop{^\sim}}}

\def\ap3m{AP$^3$M}
\def\LCDM{$\Lambda$CDM}

\def\hkpc{$h^{-1}{\ }{\rm kpc}$}
\def\hMpc{$h^{-1}{\ }{\rm Mpc}$}
\def\hMsun{$h^{-1}{\ }{\rm M_{\odot}}$}
\def\kms{${\rm{\ }km{\ }s^{-1}}$}
\def\nbody{$N$-body}
\def\c15{$c_{\rm 1/5}$}

\def\Rvir{$R_{\rm vir}$}
\def\Mvir{$M_{\rm vir}$}

\def\zform{$z_{\rm form}$}

\newcommand{\Table}[1]{Table~\ref{#1}}
\newcommand{\Sec}[1]{Section~\ref{#1}}

\newcommand{\Fig}[1]{Figure~\ref{#1}}
\newcommand{\mlapm}{\texttt{MLAPM}}
\newcommand{\mhf}{\texttt{MHF}}

\def\ea{et~al.~}                            % \ea      =  et al.
\def\lesssim{\mathrel{\hbox{\rlap{\hbox{\lower4pt\hbox{$\sim$}}}\hbox{$<$}}}}
\def\gtrsim{\mathrel{\hbox{\rlap{\hbox{\lower4pt\hbox{$\sim$}}}\hbox{$>$}}}}

\begin{document}

%%%%%%%%%%% Emulate PASA style %%%%%%%%%%%%%%%%%5
\small
\shorttitle{Interactions of Satellite Galaxies}
\shortauthor{A. Knebe~\ea}
%%%%%%%%%%% End Emulate PASA style %%%%%%%%%%%%%%%%%5

%
% Title
% Capitalise the title normally - do not use ALL CAPS.
%
\title
%%%%%%%%%%% Emulate PASA style %%%%%%%%%%%%%%%%%5
{\large \bf
%%%%%%%%%%% Emulate PASA style %%%%%%%%%%%%%%%%%5
%{
Interactions of Satellite Galaxies in Cosmological Dark Matter Halos
}
%

% Authors
% Here comes the author(s) of the paper. Please add the appropriate author
% names for your paper and indicate within the $^...$ the number(s)
% which corresponds to the institute(s) of each author. In this example
% the second author has two institutional affiliations.
% Add or remove authors as required.
% **** IMPORTANT: Leave the closing curly bracket line as is. ******

%%%%%%%%%%% Emulate PASA style %%%%%%%%%%%%%%%%%5
\author{\small 
%%%%%%%%%%% Emulate PASA style %%%%%%%%%%%%%%%%%5
%\author{
 Alexander Knebe $^{1}$, Stuart P.D. Gill, Brad K. Gibson
} % IMPORTANT: leave this curly bracket as the first character of this line.

% Date - leave this blank.
\date{}
%%%%%%%%%%% Emulate PASA style %%%%%%%%%%%%%%%%%
\twocolumn[
%%%%%%%%%%% Emulate PASA style %%%%%%%%%%%%%%%%%5
\maketitle
\vspace{-20pt}
\small
% Institutions
% Here fill in your institute name(s) and address(es)
% The number in $^...$ indicates the author number.  For example
{\center
Centre for Astrophysics \& Supercomputing, 
Swinburne University, Mail \#31, P.O. Box 218, Hawthorn, VIC 3122, 
Australia\\
$^1$aknebe@astro.swin.edu.au\\[3mm]
}

% Abstract
% Simply place your abstract between the \begin{abstract} and
% \end{abstract} commands.
%
%\begin{abstract}
%%%%%%%%%%% Emulate PASA style %%%%%%%%%%%%%%%%%
\begin{center}
{\bfseries Abstract}
\end{center}
\begin{quotation}
\begin{small}
\vspace{-5pt}
%%%%%%%%%%% Emulate PASA style %%%%%%%%%%%%%%%%%
% Place the abstract here.
We present a statistical analysis of the interactions between
satellite galaxies in cosmological dark matter halos taken from fully
self-consistent high-resolution simulations of galaxy clusters. We
show that the number distribution of satellite encounters has a tail
that extends to as many as 3--4 encounters per orbit. On average 30\%
of the substructure population had at least one encounter (per orbit)
with another satellite galaxy. However, this result depends on the age
of the dark matter host halo with a clear trend for more interactions
in younger systems.  We also report a correlation between the number
of encounters and the distance of the satellites to the centre of the
cluster: satellite galaxies closer to the centre experience more
interactions. However, this can be simply explained by the radial
distribution of the substructure population and merely reflects the
fact that the density of satellites is higher in those regions.

In order to find substructure galaxies we applied (and present) a new
technique based upon the \nbody\ code \mlapm. This new halo finder
\mhf\ (\mlapm's-halo-finder) acts with exactly the same accuracy as
the \nbody\ code itself and is therefore free of any bias and spurious
mismatch between simulation data and halo finding precision related to
numerical effects.

%%%%%%%%%%% Emulate PASA style %%%%%%%%%%%%%%%%%
%\end{abstract}
%%%%%%%%%%% End Emulate PASA style %%%%%%%%%%%%%%%%%
\end{small}
\end{quotation}
]
%%%%%%%%%%% End Emulate PASA style %%%%%%%%%%%%%%%%%

% Place keywords here. Please write all keywords in lower case. PASA uses the
% standard list of subject 
% headings adopted by The Astrophysical Journal and available from URL:
%   http://www.journals.uchicago.edu/ApJ/keywords_text.html

% A formatting command to add space between the author list and the body
% of the paper when printed. This spacing may be changed as desired.
\bigskip

%%%%%%%%%%%%%%%%%%%%%%%%%%%%%%%%%%%%%%%%%%%%%%%%%
\section{Introduction}
%%%%%%%%%%%%%%%%%%%%%%%%%%%%%%%%%%%%%%%%%%%%%%%%%
\subsubsection*{Observations}
%%%%%%%%%%%%%%%%%%%%%%%%%%%%%%%%%%%%%%%%%%%%%%%%%%%%%%%
There are several hints indicating that satellite galaxies orbiting
within our own Milky Way are interacting with each other.  Zhao
(1998), for instance, proposed a scenario where the Sagittarius Dwarf
galaxy had an encounter with the Magellanic Cloud system some 2--3
Gyrs ago, something that has also been speculated and noted by
Ibata~\& Lewis (1998).  Moreover, the two Magellanic Clouds themselves
are another example of an interacting pair of substructure
galaxies. It has also been noted by Moore~\ea (1996) that ``galaxy
harrasment'' in cosmological simulations of galaxy cluster evolution
will lead to a morphology change of satellite galaxies.

However, the literature to date lacks a statistical analysis of
interacting satellite galaxies orbiting within the potential of a
common dark matter host halo. How frequent are satellite-satellite
encounters and where in the galaxy cluster do they happen?
Furthermore, observations of the Local Group Dwarfs indicate a clear
correlation between star formation activity and the distance of the
respective Dwarf to the centre of the Milky Way (van den Bergh 1994)
with satellites farther away showing stronger activity. Can this be
ascribed to satellite-satellite interactions? The aim of this study is
to quantify such interactions in galaxy clusters derived from fully
self-consistent cosmological \nbody\ simulations within the framework
of the currently accepted Cold Dark Matter (CDM) structure formation
scenario.

\subsubsection*{Is Cold Dark Matter still feasible?}
%%%%%%%%%%%%%%%%%%%%%%%%%%%%%%%%%%%%%%%%%%%%%%%%
There is mounting, if not overwhelming, evidence that CDM provides the
most accurate description of our Universe. Observations point towards
a \LCDM\ Universe comprised of 28\% dark matter, 68\% dark energy, and
luminous baryonic matter (i.e. galaxies, stars, gas, and dust) at a
mere 4\% (cf. Spergel~\ea 2003). This so-called ``concordance model''
induces hierarchical structure formation whereby small objects form
first and subsequently merge to form progressively larger objects
(e.g. White \& Rees 1978; Davis \ea 1985).  Hence, galaxies and galaxy
clusters are constantly fed by an accretion stream of smaller entities
starting to orbit within the encompassing dark matter potential of the
host. While generally successful, the \LCDM\ model does face several
problems, one such problem actually being the prediction that
one-to-two orders of magnitude more satellite galaxies should be
orbiting within galactic halos than are actually observed (Klypin~\ea
1999; Moore~\ea 1999).

However, there are also indications that the CDM model is in fact
correct and does \textit{not} have a problem with an overabundant
population of satellite galaxies. For instance, Benson~\ea (2002)
carried out a semi-analytical study of satellites in the Local Group
and found that an earlier epoch of reionisation was sufficient to
suppress star formation in many of the subhalos and thus produce a
significant population of ``dark galaxies''. 

Therefore, if the CDM model is in fact correct and the (overabundant)
population of (dark) satellites predicted by it really does exist, it
is imperative to understand the discrepancy by investigating the
orbital evolution of these objects and their deviation from the
background dark matter distribution.

\subsubsection*{The story, so far}
%%%%%%%%%%%%%%%%%%%%%%%%%%%%%%%%%%%%%%%%%%%%%%%%
To date, typical satellite properties such as orbital parameters and
mass loss under the influence of the host halo have primarily been
investigated using \textit{static} potentials for the dark matter host 
halo (Johnston \ea 1996; Hayashi \ea 2003).  We stress that each of
these studies have provided invaluable insights into the physical
processes involved in satellite disruption; our goal is to augment
those studies by relaxing the assumption of a static host potential
as, in practice, realistic dark matter halos are neither static nor
spherically symmetric.

\subsubsection*{The story continues}
%%%%%%%%%%%%%%%%%%%%%%%%%%%%%%%%%%%%%%%%%%%%%%%%
The work presented here is based upon a set of numerical simulations
of structure formation within said concordance model, analysing in
detail the temporal and spatial properties of satellite galaxies
residing within host dark matter halos that formed fully
self-consistently within a cosmological framework. We focus on
interactions between satellite galaxies orbiting within a larger dark
matter halo and especially if there is a relation between mutual
interplay and distance to the host.  The outline of the paper is as
follows. In \Sec{Identify} we present our new halo finding algorithms
based upon the \nbody\ code \mlapm.  We then apply it to our set of
eight cosmological dark matter halos in
\Sec{Application} with a summary of ours results given in
\Sec{Summary}.

%%%%%%%%%%%%%%%%%%%%%%%%%%%%%%%%%%%%%%%%%%%%%%%%%
\section{Identifying Satellite Galaxies}\label{Identify}
%%%%%%%%%%%%%%%%%%%%%%%%%%%%%%%%%%%%%%%%%%%%%%%%%
\subsubsection*{Cosmological Simulations}
%%%%%%%%%%%%%%%%%%%%%%%%%%%%%%%%%%%%%%%%%%%%%%%%%%%
Over the last decades great advancements have been made in the
development of \nbody\ codes. We have seen the rise of tree based
gravity solvers (Barnes and Hut 1986), mesh based techniques
(Klypin~\& Shandarin 1983), and combinations of direction summation
techniques and grid based Poisson solvers (Efstathiou \ea 1985).
However, simulating the Universe in a computer and producing the data
is only the first step in a long journey; the purpose of these codes
is their predictive power, thus the ensembles of millions of dark
matter particles used with such (dissipationless) \nbody\ codes need
to be interpreted and then compared to the observable Universe. This
task requires analysis tools to map the phase-space, which is being
sampled by the particles, back to "real" objects in the Universe, the
traditional way has been through the use of "halo finders".  

\subsubsection*{Identifying Dark Matter Halos}
%%%%%%%%%%%%%%%%%%%%%%%%%%%%%%%%%%%%%%%%%%%%%%%%%%
Halo finders mine the \nbody\ data to find locally overdense
gravitationally bound systems. Under the assumption that all galaxies
and galaxies clusters are centered about local over-density peaks in
the dark matter density field they are usually found just using
spatial information of the particle distribution.  To identify objects
in this fashion, the halo finder is required in some way to reproduce
the work of the \nbody\ solver in the calculation of the density field
or the location of its peaks. The major limitation, however, will
always be the appropriate reconstruction of the density
field. Normally this task is performed \textit{after} the simulation
has finished using an independent method to derive a) the density
field and b) to smooth it on a certain scale.  With that in mind, we
are using a new method for identifying gravitationally bound objects
that utilizes the adaptive meshes of the open source \nbody\ code
\mlapm\footnote{\mlapm\ can be downloaded from the webpage
\texttt{http://astronomy.swin.edu.au/MLAPM}}(Knebe, Green~\& Binney
2001). It is called \mhf\ (\mlapm's Halo Finder) and naturally works
on-the-fly, but has also been adapted to deal with single outputs of
any \nbody\ code. However, in order to understand the functionality of
\mhf\ it is important to gain insight into the mode of operation of
\mlapm\ first.

\begin{figure}
 \begin{center}
 \psfig{file=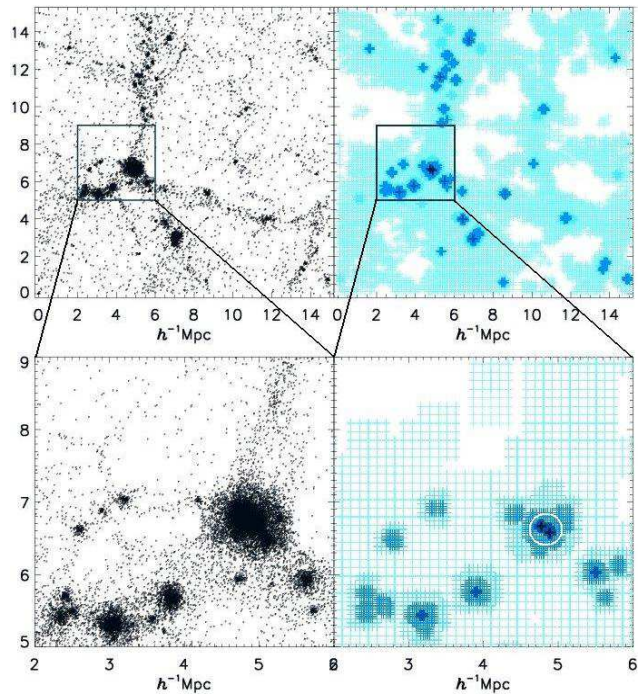,width=\hsize}
 \caption{\mlapm\ at work. The upper panels show a sample cosmological
          \LCDM\ simulation with the lower panels zooming into the
          marked region. In the left panels the particle positions are
          plotted whereas the right panels are indicating the
          (adaptive) grid points used to solve the governing equations
          of motion. The circle in the lower right panel highlights
          substructure being picked up by the finest refinement grid.}
 \label{MLAPMref}            % for cross-references
 \end{center}
\end{figure}

\begin{figure}
 \begin{center}
 \psfig{file=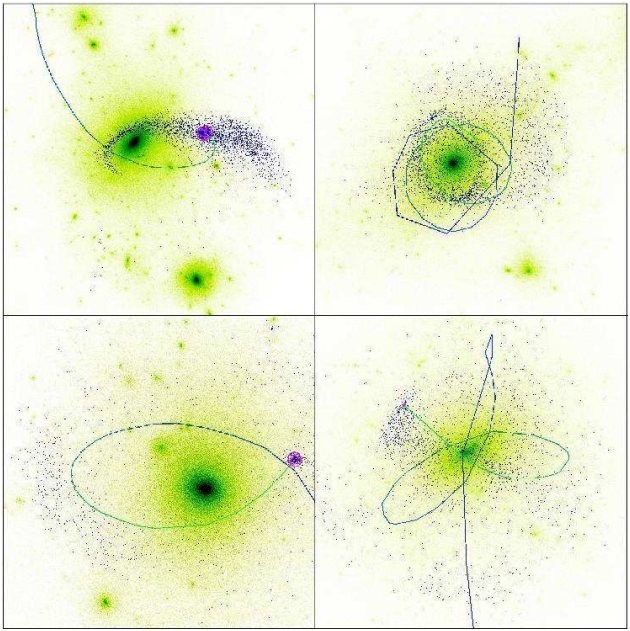,width=\hsize}
 \caption{Some sample orbits of satellite galaxies within our set
          of dark matter host halos. We can clearly see how well
          we trace the orbits and follow the tidal disruption
          of the satellites, respectively.}
 \label{Orbits}            % for cross-references
 \end{center}
\end{figure}

\subsubsection*{\mlapm's Mode of Operation}
%%%%%%%%%%%%%%%%%%%%%%%%%%%%%%%%%%%%%%%%%%%%%%%%%%%
\mlapm\ reaches high force resolution by refining high-density regions
with an automated refinement algorithm.  These adaptive meshes are
recursive: refined regions can also be refined, each subsequent
refinement having cells that are half the size of the cells in the
previous level.  This creates a hierarchy of refinement meshes of
different resolutions covering regions of interest.  The refinement is
done cell by cell (individual cells can be refined or de-refined) and
meshes are not restricted to have a particular symmetry. The criterion
for (de-)refining a cell is simply the number of particles within that
cell and a detailed study of the appropriate choice for this number
can be found elsewhere (Knebe~\ea 2001). \mlapm's adaptive refinement
meshes therefore follow the density distribution by
construction. Thus, the grid structure naturally surrounds the
(satellite) galaxies as they manifest themselves as over-densities in
the underlying background field, an example of which can be viewed in
\Fig{MLAPMref} where we show a slice through a sample \LCDM\
simulation. In the left panels the actual particle distribution is
presented whereas the right panels indicate the adaptive meshes
invoked by \mlapm\ to solve Poisson's equation and integrate the
equations of motion, respectively. In the lower right panel the white
circle highlights the ability of \mlapm'grid to locate substructure:
only on the finest refinement level it becomes apparent that the
massive galaxy cluster in fact has two centres which is a mere
reflection of the fact it recently underwent a major merger with the
two progenitors still not fully coalesced yet. The advantage of
reconstructing and using these adaptive grids to identify prospective
halo centres is that they naturally follow the density field with the
\textit{exact} accuracy of the \nbody\ code.

\subsubsection*{\mhf\ (\mlapm's-Halo-Finder)}
%%%%%%%%%%%%%%%%%%%%%%%%%%%%%%%%%%%%%%%%%%%%%%%%%%%%%%%%%%%%%
In \Fig{MLAPMref} we have seen the capability of \mlapm\ to localise
local overdensity peaks in cosmological simulations of structure
formation. But this is just the first step to identifying
gravitationally bound objects. To actually locate dark matter halos
within the simulation data we build a register of positions of the
peaks in the density field from the full adaptive grid structure
invoked by \mlapm\ using the same refinement criterion as for the
original runs; we build a list of "potential centers". To do this we
restructure the hierarchy of nested isolated \mlapm\ grids into a
"grid tree" storing the centre of the densest grid in the end of each
branch. For each of these potential centres we step out in radial bins
until the overdensity (measured in terms of the cosmological
background density) drops below the virial value set by the background
cosmological model, i.e. $\Delta_{\rm vir}=340$ for \LCDM\ at redshift
$z=0$. This defines the virial radius \Rvir\ and provides us with a
list of particles associated with that dark matter halo.

We then need to prune that list and remove (in an iterative procedure)
all gravitationally unbound particles, respectively.  Starting with
the potential centre again, we calculate the kinetic and potential
energy for each individual particle in the respective reference frame
and all particles faster than two times the escape velocity are
removed from the halo. We then recalculate the centre (as well as the
virial radius) and proceed through the process again. This iteration
stops once no further particles are removed or if there are fewer than
eight particles left in which case the potential centre will be
removed from the halo list completely. 

In the end we are left with not only a list of appropriate halo
positions but we also derived canonical properties for all credible
objects, e.g. virial radius, virial mass, velocity dispersion, density
profile, etc. A more elaborate description of our technique can be
found elsewhere though (Gill, Knebe~\& Gibson 2004a).

%%%%%%%%%%%%%%%%%%%%%%%%%%%%%%%%%%%%%%%%%%%%%%%%%
\section{Quantifying Interactions in simulated Galaxy Clusters} \label{Application}
%%%%%%%%%%%%%%%%%%%%%%%%%%%%%%%%%%%%%%%%%%%%%%%%%

\subsubsection*{The Dark Matter Host Halos}
%-------------------------------------------
We created a set of eight high-resolution galaxy clusters each
consisting of order more than a million dark matter particles. These
clusters formed in dissipationless \nbody\ simulations of the
so-called "concordance" (\LCDM) cosmology ($\Omega_0 =
0.3,\Omega_\lambda = 0.7, \Omega_b h^2 = 0.022, h = 0.7, \sigma_8 =
0.9$). The runs have a mass resolution of $m_p = 1.6 \times
10^{8}$\hMsun\ and achieved a force resolution of $\approx$2\hkpc\
allowing us to resolve the host halos down to about the central 0.25\%
of their virial radii \Rvir.

The halos were specifically selected to investigate the evolution of
satellite galaxies and its debris in an unbiased sample of host halos
thus analysing the influence of environment in the evolution of such
systems. To achieve this goal high temporal information was required
to track the development of the satellites. We therefore stored 17
outputs from $z=2.5$ to $z=0.5$ equally spaced with $\Delta t \approx
0.35$~Gyrs. From $z=0.5$ to $z=0$ we have 30 outputs spaced $\Delta t
\approx 0.17$~Gyrs.  A summary of the eight host halos is presented in
Table~\ref{HaloDetails}.

The quality of our halo finder and our data, respectively, can be
viewed in \Fig{Orbits}. There we show the orbits of four sample
satellite galaxies orbiting within their respective host halo.  This
Figure nicely demonstrates how we are very accurately tracking the
orbits of the satellites within the area of trade of the host
halos. In a companion paper (Gill~\ea 2004) we are presenting a
thorough analysis of the dynamics of these satellite galaxies. There
we also present the number distribution of orbits of the substructure
population which peaks at about 1--2 orbits with a tail extending to
as many as 5 orbits in the older systems. However, in this study we
like to focus on one particular aspect, namely satellite-satellite
encounters.

\subsubsection*{Quantifying Encounters}
%-------------------------------------------
As a first order approximation for quantifying encounters between
substructure galaxies we calculated the tidal radius of a given
satellite \textit{induced by one of the other satellites}.  This means
that the tidal radius is defined to be the radius where the
gravitational effects of the companion satellite are greater than its
self-gravity. When approximating both satellites as point masses and
maintaining that the mean density within the satellite has to be three
times the mean density of the "perturber" at distance $D$ (Jacobi
limit) the definition for tidal radius reads as follows

\begin{equation}\label{TidalRad}
 r_{\rm tidal} = \left( \frac{m}{3M} \right)^{\frac{1}{3}} D \ ,
\end{equation}

\noindent
where $m$ is the mass of the actual satellite and $M$ is the mass of
the perturbing satellite at distance $D$.

\begin{figure}
 \begin{center}
 \psfig{file=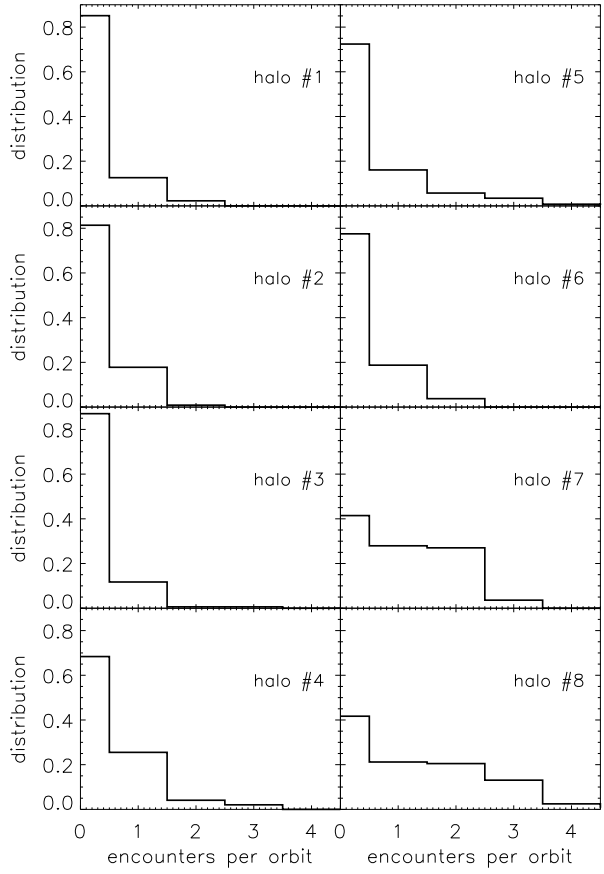,width=\hsize}
 \caption{Distribution of number of encounters for all satellite galaxies
          more massive than 10$^{10}$\hMsun\ at redshift $z=0$.}
 \label{Nenc}            % for cross-references
 \end{center}
\end{figure}

Whenever the tidal radius becomes smaller than the virial
radius\footnote{We are tracking each satellite galaxy individually
from the formation time of the host halo using its initial particle
content and hence we are in the unique position to accurately
calculate its virial radius as the radius where the mean averaged
density (measured in terms of the cosmological background density
$\rho_b$) drops below $\Delta_{\rm vir}(z)$.} of the satellite we
increased a counter for that particular satellite.  This counter now
keeps track of the number of (perturbing) interactions with companion
satellite galaxies. As some of the satellites may have had more
interactions simply because they spent more time orbiting the host we
are normalising the number of encounters by the number of orbits for
each individual satellite. The distribution of this (normalised)
counter is presented in \Fig{Nenc}. The well pronounced peak at zero
encounters shows that in most cases the interactions between
satellites is negligible.  However, we also observe that (in our
simplistic treatment for satellite-satellite interactions) we do find
as many as 3-4 encounters per orbit for individual satellites. This,
in fact, indicates that with sufficient (spatial) resolution (as it is
the case with our data) one is able to decipher the influence of the
dominant host halo from the (more minor) interactions with the
companion satellite galaxies.  We, however, leave a detailed analysis
of this phenomenon to a companion paper (Gill, Knebe~\& Gibson 2004b),
where we individually select satellite galaxies and resimulate them in
static and evolving analytic host potentials as opposed to their
evolution in the live potential used for this study.

We complement \Fig{Nenc} with \Table{Encounters} where we give the
percentage of satellites that had one or more encounters per orbit.
The average percentage amounts to 30\% of the whole substructure
population. We also observe a clear trend for the interactions to
become more prominent in younger systems. This is basically a reflection
of the fact that the younger systems are still in the process of
digesting their last major merger and have not reached an equilibrium
state yet, respectively.

\subsubsection*{Relations to Observations}
\begin{figure}
 \begin{center}
 \psfig{file=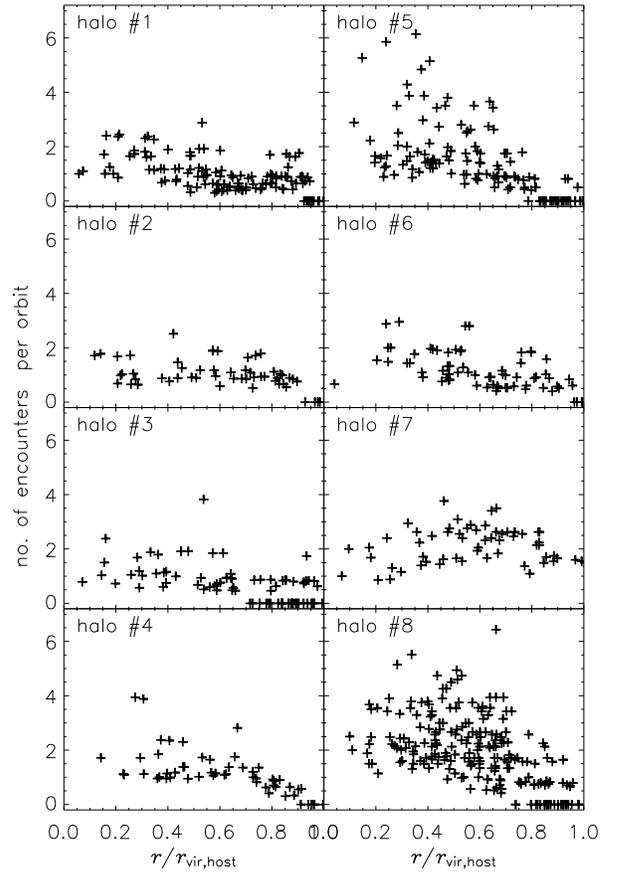,width=\hsize}
 \caption{Encounters per orbit as a function of distance to
          the host halo's centre for redshift $z=0$.}
 \label{EncDist}            % for cross-references
 \end{center}
\end{figure}

%--------------------------------------------
If we now assume that such interactions might be held responsible for
star formation bursts, i.e. if encounters trigger star formation, it
raises the question whether we can explain the observed correlation
between star formation activity in the Local Group Dwarfs and
distance to the centre of the Milky Way. Van den Bergh (1994), for
instance, reported that Dwarf spheroidals located close to the Galaxy
only experienced star formation early in their lifetimes. Dwarf
spheroidals at intermediate distances underwent significant star
formation more recently whereas the most distant ones do show ongoing
star formation at the present time. Do encounters with other
satellites trigger star formation bursts? To this extent we present
the relation between the number of encounters (per orbit) as a
function of distance to the centre of the host at redshift $z=0$. The
result can be viewed in \Fig{EncDist}.  Unfortunately we do not
observe a clear trend for all our halos, even though most of them
actually show the reverse correlation, namely the closer a satellite
to the host galaxy the more encounters with other substructure. This
relation is even more prominent when not normalising by the number of
orbits. Only halo~\#7 does show a trend that agrees with the
observational finding for star formation activity and distance to the
centre, even though we show in Gill~\ea (2004) that halo~\#7 does
otherwise have no outstanding differences to the other halos. Anyway,
as we see in Gill, Knebe~\& Gibson (2004a) the radial satellite
density distribution roughly declines like $\rho_{\rm sat} \propto
r^{-2}$ and hence the mild (anti-)correlation between number of
encounters and distance can be interpreted as a ``volume effect'':
closer to the centre of the host lives approximately the same number
of satellites in a spherical shell as farther out, but as the volume
of that shell is smaller it is more likely for the satellites to
interact.

\begin{table}
\begin{center}
\caption{Properties of the eight dark matter host halos. Distances are
         measured in \hMpc, velocities in \kms, masses in
         10$^{14}$\hMsun, and the age in Gyrs. We applied a mass-cut 
         of $M>10^{10}$\hMsun\ (100 particles) which explains the rather 
         'low' number for $N_{\rm sat}(<\!R_{\rm vir})$.}
\label{HaloDetails}
\begin{tabular}{cccccc}\hline
Halo & \Rvir &  \Mvir  & \zform & age &
$N_{\rm sat}(<\!R_{\rm vir})$ \\
 
\hline \hline
 \# 1 &  1.34 & 2.87 & 1.16 & 8.30 & 158 \\
 \# 2 &  1.06 & 1.42 & 0.96 & 7.55 &  63 \\
 \# 3 &  1.08 & 1.48 & 0.87 & 7.16 &  87 \\
 \# 4 &  0.98 & 1.10 & 0.85 & 7.07 &  57 \\
 \# 5 &  1.35 & 2.91 & 0.65 & 6.01 & 175 \\
 \# 6 &  1.05 & 1.37 & 0.65 & 6.01 &  85 \\
 \# 7 &  1.01 & 1.21 & 0.43 & 4.52 &  59 \\
 \# 8 &  1.38 & 3.08 & 0.30 & 3.42 & 251 \\
\end{tabular}
\end{center}
\end{table}

\begin{table}
\begin{center}
\caption{Percentage of satellites that had one or more
         encounters per orbit.}
\label{Encounters}
\begin{tabular}{cc}\hline
Halo & percentage \\
 
\hline \hline
 \# 1 &  14 \\
 \# 2 &  18 \\
 \# 3 &  12 \\
 \# 4 &  31 \\
 \# 5 &  27 \\
 \# 6 &  22 \\
 \# 7 &  58 \\
 \# 8 &  58 \\
\end{tabular}
\end{center}
\end{table}
%%%%%%%%%%%%%%%%%%%%%%%%%%%%%%%%%%%%%%%%%%%%%%%%%
\section{Summary} \label{Summary}
%%%%%%%%%%%%%%%%%%%%%%%%%%%%%%%%%%%%%%%%%%%%%%%%%
We used a set of eight high-resolution cosmological simulations to
investigate and quantify interactions between satellite galaxies
orbiting within a common dark matter halo. Using our definition for
encounter, which is based upon the mutually induced tidal radius, we
showed that on average 30\% of the substructure population had had
more than one encounter per orbit with another satellite galaxy
orbiting within the same host halo. There is, however, a clear trend
for interactions to be more common in young galaxy clusters.  We
furthermore showed that satellite galaxies closer to the centre of the
host halo had had more interactions with companion satellites, not
because they simply orbited for longer in the underlying host
potential but most likely because of the universal radial distribution
of satellite galaxies found in cosmological dark matter halos
(Gill~\ea 2004). Even though satellite-satellite interactions are
unimportant for the majority of satellite galaxies, there exists a
sub-population for which this needs to be investigated in more detail
and more carefully, respectively.

We also noted that there is a degeneracy between the influence of the
host halo and the interactions with the companion satellites which can
only be disentangled with an appropriate resolution for both the
actual \nbody-simulation and the halo finding technique. We therefore
applied a new method for identifying gravitationally bound objects
in cosmological \nbody\ simulations. This new technique is based upon
the adaptive grid structures of the open source adaptive mesh
refinement code \mlapm\ (Knebe, Green~\& Binney 2001). The halo finder
is called \mhf\ and acts on the same accuracy level as the actual
simulation. A more thorough study of the functionality of \mhf\ is
presented in Gill, Knebe~\& Gibson (2004a). A detailed analysis
of the degeneracy between influence of the host halo and interactions
with companion satellites can be found in a companion paper, too
(Gill, Knebe~\& Gibson 2004b).

%%%%%%%%%%%%%%%%%%%%%%%%%%%%%%%%%%%%%%%%%%%%%%%%%
\section*{Acknowledgments}
%%%%%%%%%%%%%%%%%%%%%%%%%%%%%%%%%%%%%%%%%%%%%%%%%
The simulations presented in this paper were carried out on the
Beowulf cluster at the Centre for Astrophysics \& Supercomputing,
Swinburne University.

%%%%%%%%%%%%%%%%%%%%%%%%%%%%%%%%%%%%%%%%%%%%%%%%%
\section*{References}
%%%%%%%%%%%%%%%%%%%%%%%%%%%%%%%%%%%%%%%%%%%%%%%%%

% PASA uses the same conventions as ApJ for journal abbreviations.  Sample
% references are as follows. 
% Please follow the same format for your references.

\reference Barnes J. E., Hut P., 1986, Nature 324, 446
\reference Benson A.J., Frenk C.S., Lacey C.G., Baugh C.M., Cole S., 2002, MNRAS 333, 177
\reference Bertschinger E., 1998, ARA\&A 36, 599
\reference Davis M., Efstathiou G., Frenk C. S., \& White S. D. M., 1985, ApJ 292, 371
\reference Efstathiou G., Davis M., White S.D.M., Frenk C.S., 1985, ApJS 57, 241
\reference Gill S.P.D., Knebe A., Gibson B.K., 2004, MNRAS submitted
\reference Gill S.P.D., Knebe A., Gibson B.K., 2004, in preparation
\reference Gill S.P.D., Knebe A., Gibson B.K., Dopita M.A., 2004, MNRAS submitted
\reference Hayashi E., Navarro J., Taylor J., Stadel J., Quinn T., 2003, ApJ 584, 541
\reference Ibata R.A., Lewis G.F., 1998, ApJ 500, 575
\reference Johnston K., Hernquist L., Bolte M., 1996, ApJ 465, 278
\reference Klypin A.A., Shandarin, S. F., 1983, MNRAS 204, 891
\reference Klypin A., Kravtsov A., Valenzuela O., Prada F., 1999, ApJ 522, 82
\reference Knebe A., Green A., Binney J., 2001, MNRAS 325, 845
\reference Moore B., Ghigna S., Governato F., Lake G., Quinn T., Stadel J., 
           Tozzi P., 1999, ApJL 524, 19
\reference Spergel D. N., Verde L., Peiris H. V., Komatsu E., Nolta M. R., 
           Bennett C. L., Halpern M., Hinshaw G., Jarosik N., Kogut A., 
           Limon M., Meyer S. S., Page L., Tucker G. S., Weiland J. L., 
           Wollack E., Wright E. L., 2003, ApJS 148, 175
\reference van den Bergh S., 1994, ApJ 428, 617
\reference White S. D. M., \& Rees M., 1978, MNRAS 183, 341
\reference Zhao H., 1998, ApJL 500, 149

\end{document}